\documentclass[aps,prb,final,reprint,twocolumn,floatfix,showpacs,superscriptaddress,notitlepage,longbibliography]{revtex4-2}

\usepackage[latin9]{inputenc}
\usepackage{graphicx,amssymb,soul,color,amsmath,bm}
\usepackage{epstopdf,hyperref,subfigure,braket,dsfont}
\usepackage[normalem]{ulem}

\hypersetup{colorlinks=true,citecolor=blue,linkcolor=magenta}

\sethlcolor{yellow}

\newcommand{\refeq}[1]{(\ref{#1})}
\allowdisplaybreaks

\begin{document}

\title{
Universal quasi-Fermi liquid physics of one-dimensional interacting fermions
}

\author{Joshua D. Baktay}
\affiliation{Department of Physics, Northeastern University, Boston, Massachusetts 02115, USA}

\author{Adrian E. Feiguin}
\affiliation{Department of Physics, Northeastern University, Boston, Massachusetts 02115, USA}

\author{Juli\'an Rinc\'on}
\affiliation{Department of Physics, Universidad de los Andes, Bogot\'a, D.C. 111711, Colombia}

\date{\today}

\begin{abstract}
We present a class of one-dimensional, generic, spinless fermion lattice Hamiltonians that express quasi-Fermi liquid physics, manifesting both Luttinger and Fermi liquid features solely due to irrelevant interactions. Using infinite matrix product state techniques, we unveil its universal structure by calculating static and dynamic responses. Key features include a finite discontinuity in the momentum distribution at the Fermi level, despite power-law singularities in the spectral function protected by particle-hole symmetry. Away from half filling, Landau quasiparticles emerge. Charge dynamics show either high-energy bound states or a concentration of spectral weight within the continuum for attractive or repulsive interactions, respectively. These universal features are realized across multiple models and energy scales, thus reifying the quasi-Fermi liquid as a unique paradigm for one-dimensional fermions.
\end{abstract}

\maketitle

\paragraph*{Introduction.} Exploring quantum phases is a central aim of condensed-matter physics. In particular, systems that realize complex behavior that cannot be explained simply in terms of the individual, noninteracting degrees of freedom, are of great theoretical interest. One-dimensional interacting fermions provide a prime example in which the physics cannot be described in terms of individual fermionic quasiparticles. Due to the pervasive nesting, present at all electronic densities, excitations are instead collective modes, and the low-energy physics is understood in the context of Luttinger liquid (LL)~\cite{Haldane_1981} theory. Collective modes can generally be interpreted as some form of density fluctuations, which in turn are naturally expressed in terms of bosonic degrees of freedom, instead of fermionic. This contrasts significantly with Fermi liquid theory, the prevailing paradigm to describe the problem in higher dimensions. In Fermi liquid theory (FL), excitations are fermionic quasiparticles carrying the same quantum numbers as the original electrons, and the spectrum near the Fermi surface resembles that of noninteracting systems, with sharply defined coherent excitations in the form of delta-like peaks with a finite lifetime. Instead, LLs reveal the absence of Landau quasiparticles, manifesting only power-law singularities in the single-particle spectrum. Such phenomena are universal; hence realized within a wide range of systems, regardless of microscopic details. In this Letter we focus on a recently found regime of one-dimensional (1D) physics known as quasi-Fermi liquids (qFLs), that manifests signatures of both LL and FL physics~\cite{Rozhkov_2006, Rozhkov_2014, Baktay_2023}.

Conventional LL theory assumes that the noninteracting dispersion is linear, hence ignoring the effects of the lattice with its dispersion $\epsilon(k)=-2t\cos{k}$. Physics beyond the LL paradigm has focused on reintroducing the curvature of the dispersion relation [known as nonlinear LL (nLL)]~\cite{Pustilnik_2006, Khodas_2007, Pereira_2007, Periera_2012, Imambekov_2012}, with experimental tests~\cite{Barak2010, Jin2019, Wang2020}. Other ideas focused on FL features explored the multiparticle excitations of 1D integrable models~\cite{Carmelo_1991a, Carmelo_1991b, Carmelo_1992}. In contrast, the qFL contains \emph{only} irrelevant interactions in the renormalization group sense, in addition to the nonlinear dispersion~\cite{Rozhkov_2006, Rozhkov_2014}. Therefore, since fundamental properties are generally defined by the asymptotic behavior of propagators, nLL and qFL occupy different universality classes: nLL are perturbatively connected to LL, while qFL to noninteracting fermions, \emph{\`a la} FL.

Here, we present a class of microscopic fermion lattice Hamiltonians that express qFL features across various local interactions. We numerically study the ground state directly in the thermodynamic limit using uniform matrix product states (uMPS) and the variational uMPS (VUMPS) algorithm~\cite{Zauner-Stauber_2018, Vanderstraeten_2019}. We study excitations, namely the spectral function (SF) and dynamic structure factor (DSF), using tangent-space MPS methods~\cite{Haegeman_2012, Haegeman_2013, Vanderstraeten_2019}. Our results are further refined using finite-entanglement scaling (FES)~\cite{Tagliacozzo_2008, Pollman_2009, Pirvu_2012, Stojevic_2015, Rams_2018, Vanhecke_2019}. By these means, we characterize the qFL in a generic class of models, reifying its robustness as a paradigm of 1D physics.

\paragraph*{Models and methods.\label{sec:methods}} We construct a class of interacting spinless fermion models on a 1D infinite lattice (defined by points $x \in \mathbb Z$), with general interactions over three contiguous lattice sites, and study their critical regime. The Hamiltonian characterizing this class is
\begin{equation}
    H = -t\sum_{x \in \mathbb{Z}} ( c_x^\dagger c_{x + 1} + \textrm{H.c.} ) - \mu N + H^{\rm int},
    \label{eq:Hclass}
\end{equation}
where the hopping $(t)$ and chemical potential $(\mu)$ terms define the noninteracting part of $H$, setting the bandwidth and particle density, apiece. The interaction $H^{\rm int}$, always composed of two interacting three-site terms, can take any of the following forms:
\begin{align}
    H_{VV_2}^{\rm int} &= V\sum_{x \in \mathbb{Z}} n_x n_{x + 1} + V_2\sum_{x \in \mathbb{Z}} n_x n_{x + 2} 
    \label{eq:VV2} \\
    H_{Vt_c}^{\rm int} &= V \sum_{x \in \mathbb{Z}} n_x n_{x + 1} + t_c \sum_{x \in \mathbb{Z}} n_x (c_{x + 1}^\dagger c_{x + 2} + \textrm{H.c.} ) 
    \label{eq:Vtc} \\
    H_{Vt_c'}^{\rm int} &= V \sum_{x \in \mathbb{Z}} n_x n_{x + 1} + t'_c \sum_{x \in \mathbb{Z}} ( c_x^\dagger n_{x + 1} c_{x + 2} + \textrm{H.c.} ) 
    \label{eq:Vtc2} \\
    H_{V_2t_c}^{\rm int} &= V_2 \sum_{x \in \mathbb{Z}} n_x n_{x + 2} + t_c \sum_{x \in \mathbb{Z}} n_x ( c_{x + 1}^\dagger c_{x + 2} + \textrm{H.c.} )
    \label{eq:V2tc}
\end{align}
%
%
In other words, we study four different Hamiltonians defined by their interacting part, which we shall refer to by their couplings, as shown in \refeq{eq:VV2}--\refeq{eq:V2tc}: $VV_2$ at half filling and $Vt_c$, $Vt'_c$, $V_2t_c$ away from half filling~\cite{footnote3}. The $VV_2$ model away from half filling was explored in Ref.~\cite{Baktay_2023}. The parameters $V$ and $V_2$ denote nearest- and next-nearest-neighbor interactions, respectively, and $t_c$ and $t'_c$ correspond to correlated-hopping amplitudes, with $n_x = c^\dagger_x c_x - 1/2$. Energies are expressed in units of $t$. We note that these are all possible charge-conserving interactions in a spinless next-nearest-neighbor model with two interacting terms: This is the defining characteristic of our Hamiltonian class.

To characterize the ground state and excitations of $H$ we utilized uMPS, a type of MPS that manifestly encodes translational invariance, and tangent-space methods, which exploit the variations of an MPS on its tangent plane. Both work directly in the thermodynamic limit~\cite{Vidal_2007, Haegeman_2011, Vanderstraeten_2019, Cirac_2021}. The representation power of these variational ansatzes (how close they approximate arbitrary many-body states) is determined by their bond dimension $\chi$ that controls the number of variational parameters and defines a manifold as a subspace of the (exponentially large) many-body Hilbert space. VUMPS variationally optimizes a uMPS to approximate the ground state within such a manifold~\cite{Zauner-Stauber_2018, Vanderstraeten_2019}. The accuracy of the optimized uMPS is characterized by the energy-density error, the discarded weight, and the tangent-vector norm, which our simulations achieve around $10^{-12}$, $10^{-10}$, and $10^{-13}$, respectively. To study excitations we use the so-called quasiparticle ansatz~\cite{Haegeman_2012, Haegeman_2013, Vanderstraeten_2019}, which approximates excitations as well-defined momentum superpositions of local perturbations above the uMPS ground state. This ultimately amounts to a momentum-dependent eigenvalue problem whose energy spectrum scales with $\chi$.
We carried out simulations with bond dimensions up to $\chi=640$ for ground state and $\chi=192$ for excitations, keeping the entire spectrum allowed by $\chi$. We note that tensor network methods are nonperturbative \footnote{These methods are limited by the amount of entanglement they can capture; however, this can be overcome using FES. Tangent-space methods such as the quasiparticle ansatz cannot describe multiparticle excitations or continua and are limited to low-lying excitations.}.

Finite $\chi$ imposes truncation effects on all calculations. We address this using FES~\cite{Tagliacozzo_2008, Pollman_2009, Pirvu_2012, Stojevic_2015, Rams_2018, Vanhecke_2019}, which allows to further refine the results. By studying the behavior of expectation values as a function of $\chi$, we can extrapolate to $\chi \to \infty$ which corresponds to the exact result. This FES analysis was done for all relevant calculations whose details are expounded upon below.

\paragraph*{Results.\label{sec:results}} For generic locally-interacting fermions, the qFL can be realized by setting the couplings of $H$ such that marginal terms in the underlying effective field theory do not contribute, leaving the irrelevant terms to stabilize the state. We accomplished this by exploring the interplay between the terms that define $H^{\rm int}$ in \refeq{eq:Hclass}, for each of the four combinations of the Hamiltonian class. The field theory is equivalent to that studied in Ref.~\cite{Rozhkov_2014}, which omits marginal interactions \emph{a priori}.

To this end, we first analyze the ground state of the class of models defined by $H$, following the recipe outlined in Ref.~\cite{Baktay_2023}. By judiciously searching the Hamiltonian parameter space of all models, we found continuous regions where the Luttinger parameter $K \simeq 1$~\cite{suppmat}, which measures the degree of correlations. Of these regions, we find that all models exhibit a finite discontinuity in the momentum distribution at the Fermi momentum, $k_F$, which is a FL hallmark. This is demonstrated by performing a FES study on the discontinuity size. This stands in contrast to LL which manifest a power-law singularity. 
In what follows, we have selected representative sets of parameters to study their excitations: $(V, V_2, n) = (-1.2, 1.2, 0.5)$, $(V, t_c, n) = (-1.9, 0.8, 0.25)$, $(V, t'_c, n) = (1.4, 0.8, 0.41)$, $(V_2, t_c, n) = (-1, 0.8, 0.26)$.

The discontinuity in the momentum distribution for $K\simeq1$ shows that the qFL ground state is perturbatively connected to free fermions. This is sensible given the low-energy asymptotics of the field theory: Marginal terms have been nullified and irrelevant interactions vanish by definition for $k = k_F$.


\begin{figure}
    \centering
    \includegraphics*[width=.48\textwidth]{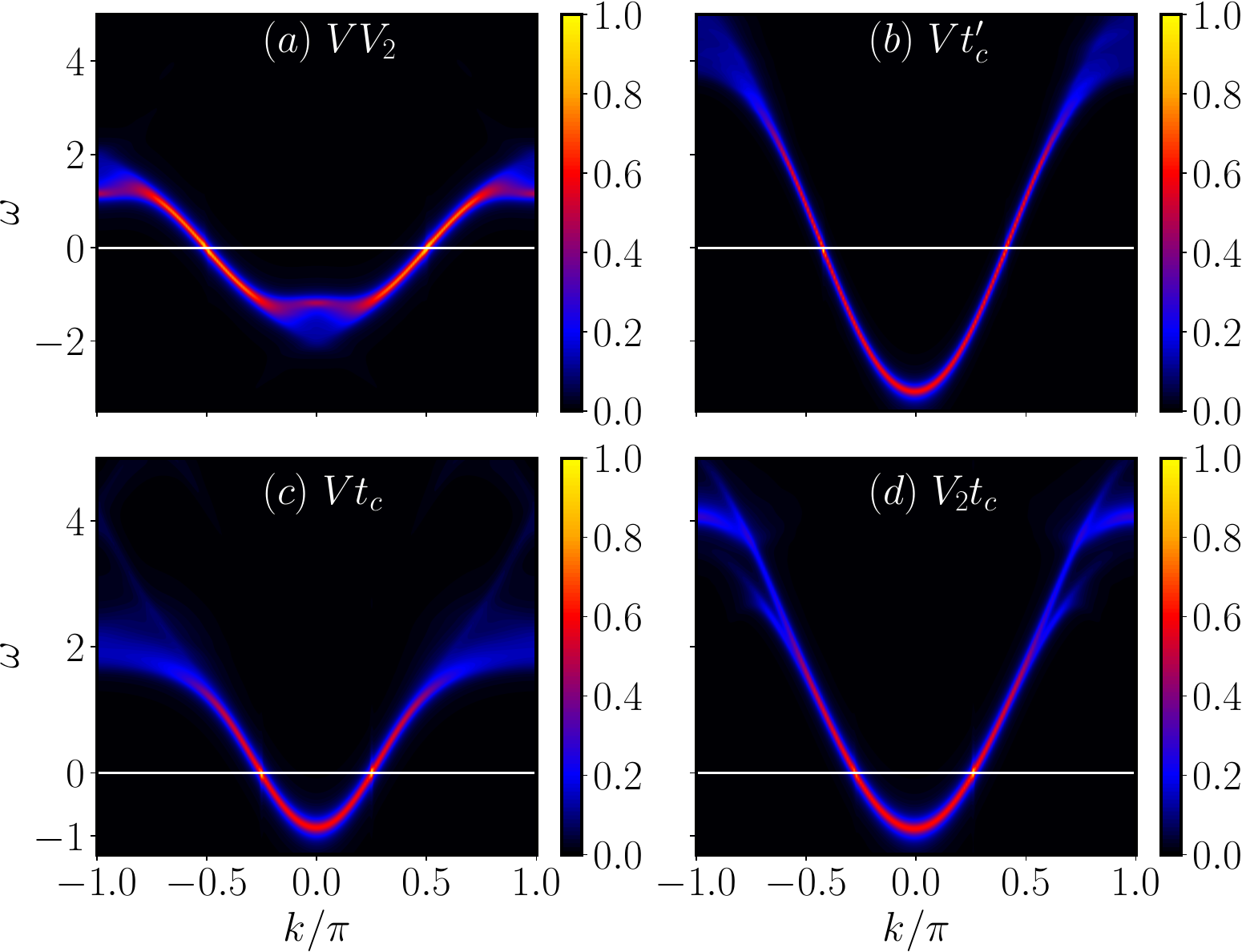}
    \caption{Normalized spectral functions for four qFL models and their densities ($\chi=104$ with full excitation spectrum): (a) $VV_2$, $n = 0.5$; (b) $Vt'_c$, $n = 0.41$; (c) $Vt_c$, $n = 0.25$; (d) $V_2t_c$, $n = 0.26$. The horizontal line indicates the Fermi level. Notice the particle-hole symmetry for the $VV_2$ model relative to the asymmetry of the others.
    }
    \label{fig:Ak}
\end{figure}

\paragraph*{Spectral function.} Single-particle excitations can be studied via the SF, $A(k,\omega) = A_p(k, \omega) + A_h(k, \omega)$,
\begin{equation}
A_{p/h}(k, \omega) = \sum\nolimits_\alpha |\langle \alpha|c^{(\dagger)}_k|0\rangle|^2\delta(\omega \mp (E_\alpha-E_0)),
\label{Akw}
\end{equation}
which include particle ($p$) and hole ($h$) contributions. $|0\rangle$ is the ground state with energy $E_0$, $|\alpha\rangle$ are excited states with one extra particle/hole with energy $E_\alpha$, and $c^{(\dagger)}_k$ creates a particle/hole with momentum $k$. In noninteracting systems $A(k,\omega)$ consists of delta peaks broadened into a Lorentzian in the presence of interactions for dimensions larger than one, according to FL (in addition to an incoherent high-energy background). In contrast, according to LL, in 1D the interactions cause power-law singularities~\cite{Dzyaloshinskii_1974, Luther_1974, Meden_1992, Voit_1993, Giamarchi, Gogolinbook}.

Modifications beyond the linearized LL reintroduce curvature dispersion, which is irrelevant in the renormalization group sense, alongside marginal interactions~\cite{Imambekov_2009a, Imambekov_2009b, Pereira_2009, Pustilnik_2006, Khodas_2007, Imambekov_2012}. Since the qFL contains \emph{only} irrelevant interactions alongside the nonlinear dispersion, its excitations retain their quasiparticle nature as $k \rightarrow k_F$ instead of returning to the LL power-law~\cite{Khodas_2007}.

At finite momentum, behavior in the particle and hole sectors varies depending on the quasiparticle group velocity $v$, relative to the Fermi velocity $v_F$~\cite{Rozhkov_2014}. For $v > v_F$, scattering is dispersive giving excitations a finite lifetime and capping the spectral weight near the single-particle energy yielding Lorentzian-like lineshapes. This is possible because the irrelevant curvature allows the dispersion to now lie within the continuum of support~\cite{Khodas_2007}. For $v < v_F$, scattering is not dispersive and causes proliferation of low-energy particle-hole pairs. Such behavior leads to an orthogonality catastrophe characteristic of LL and manifests as edge singularities accompanied by power laws in $A(k,\omega)$~\cite{Giamarchi, Gogolinbook, Pustilnik_2006, Rozhkov_2014}. Thus, we summarize the hallmarks of the qFL as the presence of power-law singularities in \emph{at least} one sector of the SF (\emph{despite} the momentum distribution finite discontinuity) with potential for quasiparticles in the opposite sector.

\begin{figure}
    \centering
    \includegraphics*[width=.35\textwidth]{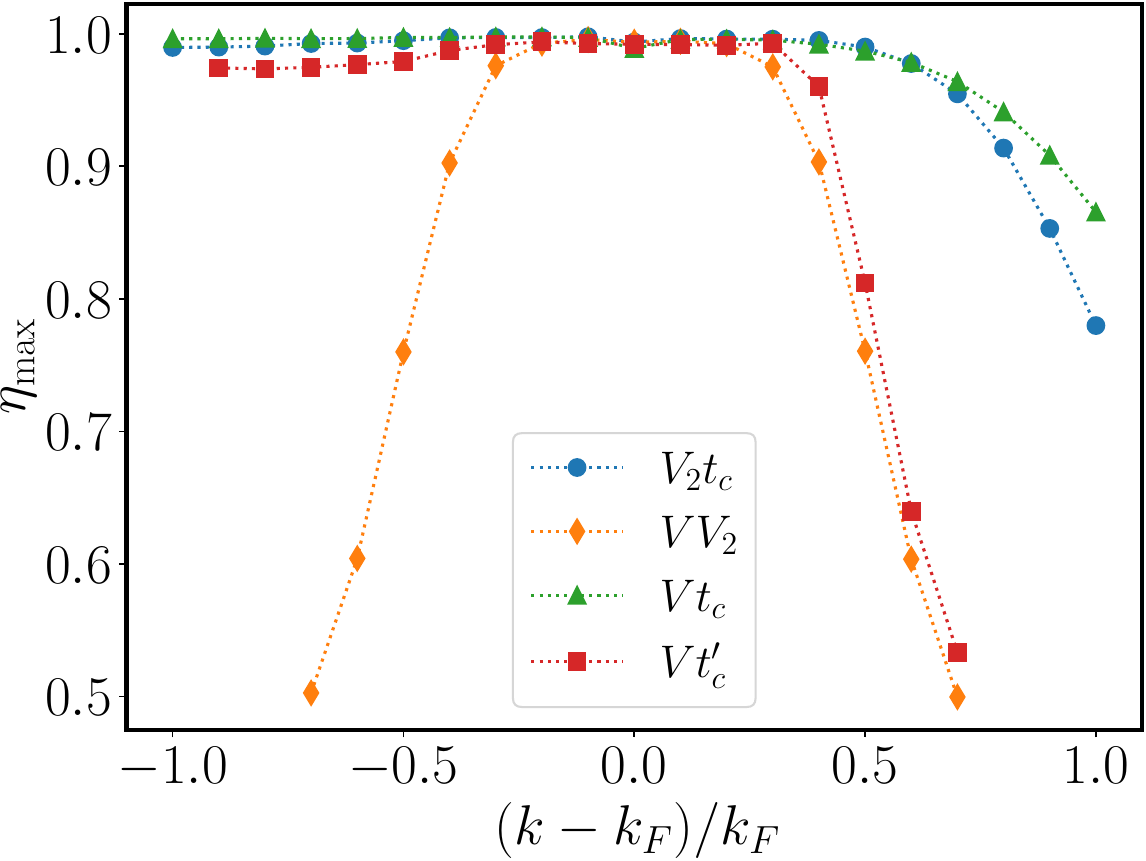}
    \caption{FES scaling exponents as a function of momentum for the qFL models shown in Fig.~\ref{fig:Ak}. $\eta_{\max} = 1~(\eta_{\max} < 1)$ corresponds to quasiparticle (power-law) excitations.
    }
    \label{fig:etak}
\end{figure}

To show this for our class of Hamiltonians, we computed the SF using the quasiparticle ansatz~\cite{Haegeman_2012, Haegeman_2013, Vanderstraeten_2019, suppmat}. 
The spectrum becomes continuous as $\chi \rightarrow \infty$. At finite $\chi$ it is discrete which requires the delta function of Eq.~\refeq{Akw} to be replaced by Lorentzians with artificial broadening $\gamma$. So long as $\gamma$ is larger than the level spacing we can mitigate finite-entanglement (finite-$\chi$) effects. In the scaling limits $\chi \rightarrow \infty$, $\gamma \rightarrow 0$, we recover Eq.~\refeq{Akw}.

Our results for the SF can be seen in Fig.~\ref{fig:Ak}. The horizontal line depicts the Fermi energy with hole excitations for $\omega < 0, |k| < k_F$ and particle excitations for $\omega > 0, |k| > k_F$. The SF is best characterized by considering fixed values of $k$ (momentum cuts), above and below $k_F$, as we now discuss. The qFL signatures described above can be seen in Fig.~\ref{fig:etak} depicting scaling exponents, $\eta_{\max}$, as a function of momentum for the qFL class. 
Inspired by Refs.~\cite{Jeckelmann_2002, Benthien_2004}, we can extract $\eta_{\max}$ from momentum cuts by extending FES for dynamic correlations~\cite{suppmat}, allowing us to track how the peak-maximum spectral weight scales with $\xi$ and $\gamma$. The correlation length of the ground-state uMPS, $\xi$, depends on $\chi$. 
First performing the $\gamma$ scaling to extract $\eta_{\max}$ for a series of fixed $\xi$, we then extrapolate to $\xi \rightarrow \infty$ by fitting $\eta_{\max}$ against $\xi^{-1}(\chi)$. The obtained exponents characterize the nature of the excitations of the particle and hole sectors: $0 < \eta_{\max} < 1$ indicates power laws whereas $\eta_{\max} \simeq 1$ specifies quasiparticles.

In three of the four models~\refeq{eq:Hclass} we see the \emph{mixed} qFL signature described above: quasiparticles in the hole sector and power laws in the particle sector. Away from half filling, the asymmetry between the sectors is determined by the group velocity of the quasiparticles, $v$. For generic dispersions with density $n < 1 / 2$ one typically expects $v < v_F$ for a hole. Thus, one might always expect power-law edge singularities in the hole sector. However, approximating the dispersion to second order can yield, in the strongly interacting regime, an effective mass with sign opposite to $v_F$~\cite{Pereira_2009, Pereira_2007}, flipping the sign of $v$ relative to $v_F$. Note, additionally, one can swap the behavior of the sectors under a particle-hole transformation. In the $VV_2$ model we realize the \emph{full} qFL signature as power laws in both sectors due to protection of particle-hole symmetry at half filling. In this case, the dispersion coincides with the edge of the spectral support in both sectors~\cite{Khodas_2007}.


For all models~\refeq{eq:Hclass} within our class of Hamiltonians, the decrease in $\eta_{\max}$ in the power-law sectors is due to the momentum dependence of the couplings within the effective field theory~\cite{Rozhkov_2014}. As $k$ increases the coupling increases along with the power-law severity. As $k \rightarrow k_F$ the coupling goes to zero, the irrelevant terms vanish, and $\eta_{\max} \rightarrow 1$. We confirm this asymptotic behavior by computing $\eta_{\max}$ directly at $k=k_F$ which demonstrates the universality of the qFL.

\begin{figure}
    \centering
    \includegraphics*[width=.48\textwidth]{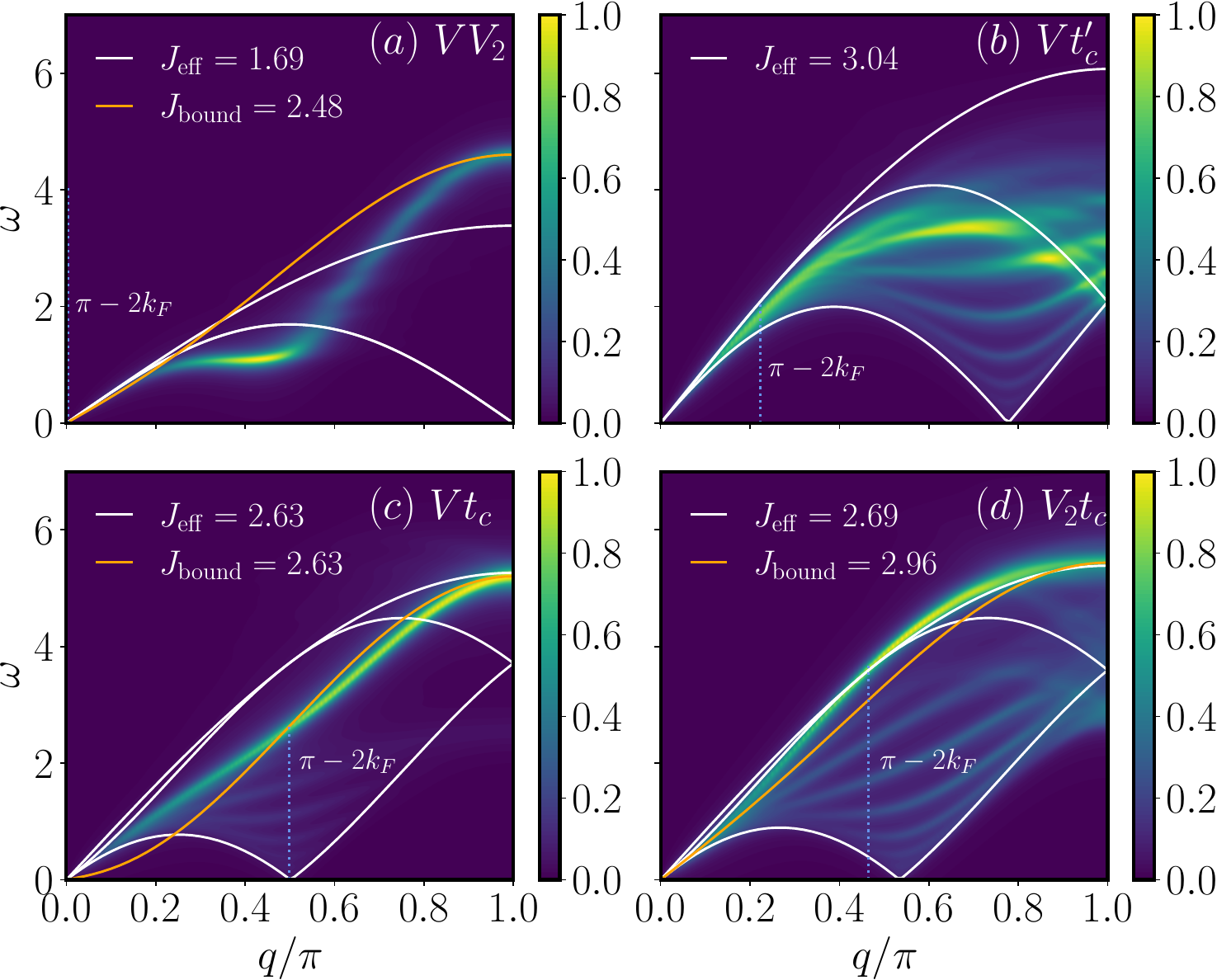}
    \caption{Normalized dynamic structure factor for four qFL models ($\chi=104$ with full excitation spectrum): (a) $VV_2$, (b) $Vt'_c$, (c) $Vt_c$, (d) $V_2t_c$. There are two branches at half-filling (a) and three otherwise (b)-(d). Note the dominant upper branch in (a),(c),(d) above a faint continuum [see Figs.~\ref{fig:etaq}(a) and \ref{fig:etaq}(b)], mimicking the {\it attractive}-$tV$ model, suggesting bound states. Next, note the dominant middle branch in (b) along with faint upper/lower branches, mimicking the {\it repulsive}-$tV$ model. Strengthening the comparison, we plot the free-fermion dispersion (white), using a bandwidth $J_{\rm eff}$ calculated from $A(k,\omega)$, and the ferromagnetic-XXZ magnon dispersion (orange); $J_{\rm bound}$ is fit to the energy scale.
    The discrete lines will fuse into a smooth continuum as $\chi \to \infty$.
    }
    \label{fig:dsf}
\end{figure}

\paragraph*{Dynamic structure factor.} We compute the DSF as the density-density dynamic correlation function, 
\begin{equation}
S(q, \omega)= \sum\nolimits_\alpha |\langle \alpha|\rho_q|0\rangle|^2\delta(\omega-(E_\alpha - E_0)),
\label{dsf}
\end{equation}
where $\rho_q$ creates a particle-hole excitation of momentum $q$ on top of the ground state. Again, at finite $\chi$ the deltas are replaced by Lorentzians with broadening $\gamma$. This allows for smooth momentum cuts with oscillations that reflect the artificially discrete spectrum due to finite $\chi$. The DSF for the qFL class is shown in Fig.~\ref{fig:dsf}, from which universal features can be extracted at length scales differentiated by $q \sim \pi - 2k_F$ as an approximate switch between the two- and three-branch regimes (away from half filling). The branches correspond to thresholds that define the support of the two-particle spectrum~\cite{Pereira_2009}.

For $q \lesssim \pi - 2k_F$ we observe the expected quadratic broadening of the linear sound mode for all models due to the nonlinear dispersion and irrelevant interactions~\cite{Rozhkov_2005, Pereira_2007}. In addition, DSF spectral weight is concentrated on the upper threshold of the spectrum, similar to free fermions~\cite{suppmat}. This demonstrates the FL-like nature of the qFL low-energy density excitations.

For $q \gtrsim \pi - 2k_F$, the absence of marginal terms at low energies breaks down and the larger parameter in absolute value, within $H^{\rm int}$, dominates whose positive (negative) sign causes a net-repulsive (attractive) behavior. This is most apparent through the behavior of the branches. For net-attractive models with negative parameter [$VV_2$, $Vt_c$, $V_2t_c$, Figs.~\ref{fig:dsf}(a)--\ref{fig:dsf}(d)], spectral weight is concentrated in the upper branch, reminiscent of the bound state/magnon dispersion of the attractive-$tV$/ferromagnetic-XXZ model~\cite{Viswanath_1995, Franchini, Pereira_2008}. For net-repulsive models with the positive parameter [$Vt'_c$, Fig.~\ref{fig:dsf}(b)], spectral weight is instead concentrated in the middle branch, reminiscent of the repulsive-$tV$ model~\cite{Caux_2005a, Caux_2005b, Pereira_2009}.

\begin{figure}
  \centering
  \includegraphics*[width=0.24\textwidth]{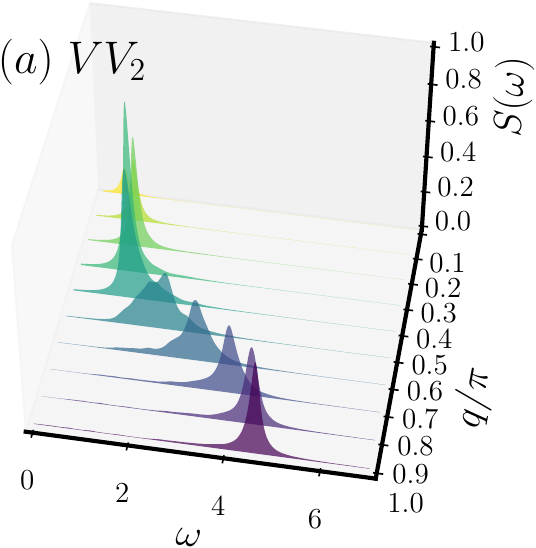}
  \raisebox{1.5em}{\includegraphics*[width=0.23\textwidth]{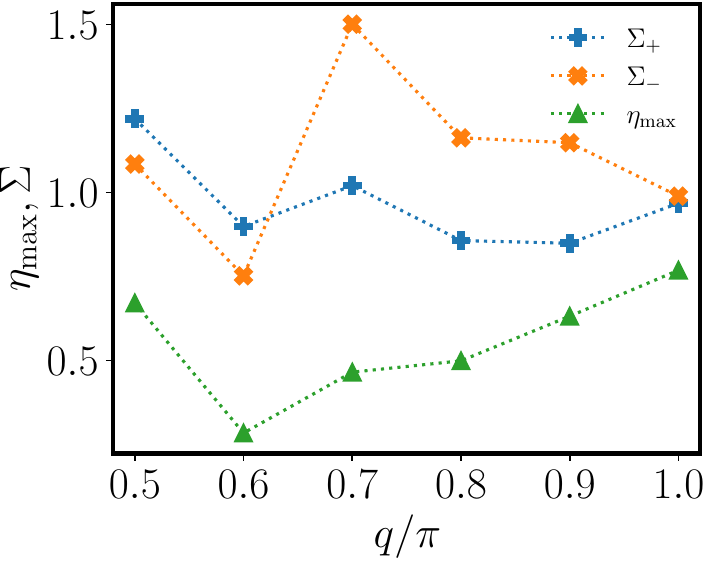}}
  \includegraphics*[width=0.24\textwidth]{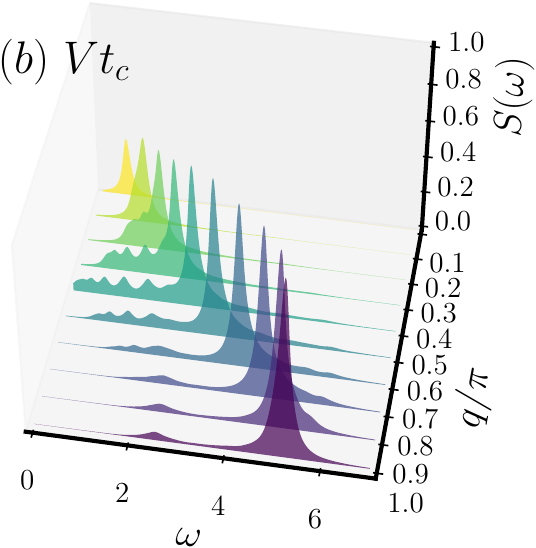}
  \raisebox{1.5em}{\includegraphics*[width=0.23\textwidth]{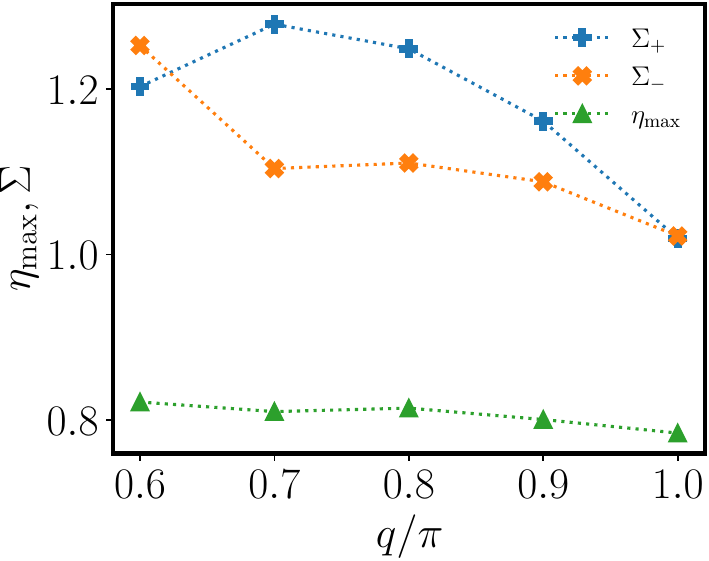}}
  \includegraphics*[width=0.24\textwidth]{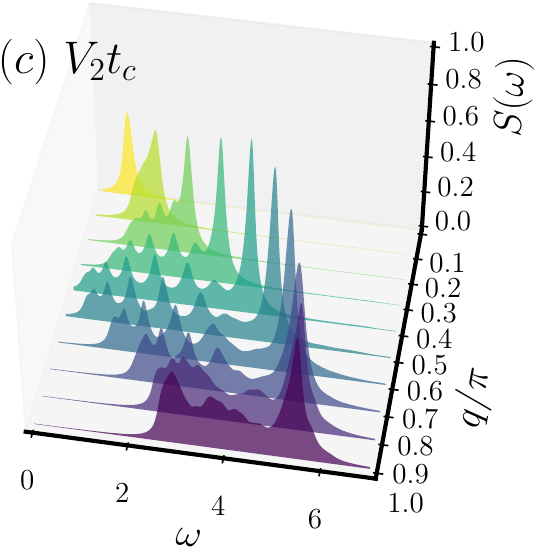}
  \raisebox{1.5em}{\includegraphics*[width=0.23\textwidth]{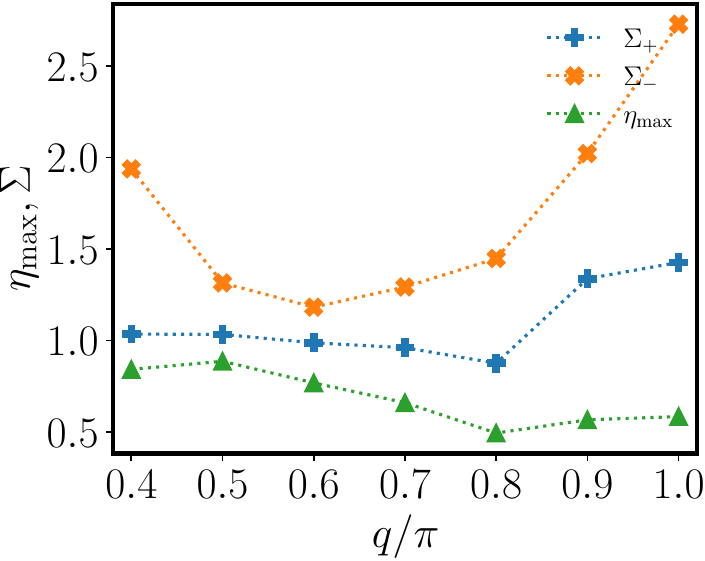}}
  \caption{FES scaling exponents, $\eta_{\max}$, and half width at half maximum, $\Sigma_\pm$, as a function of momentum (right column) of the peaks in the momentum cuts (left column) for the three attractive-qFL models. 
  For (a) and (b), as $q \rightarrow \pi$ the lineshape evolves toward a Lorentzian-like peak ($\Sigma \rightarrow 1$). In (c) we have a similar behavior instead near $q=0.6\pi$. Given effective attractive interactions, this peak represents an exciton hybridized with the surrounding excitations causing $\eta_{\rm max} < 1$.
    } 
  \label{fig:etaq}
\end{figure}

To make this notion more rigorous, we plot free-fermion and magnon dispersions on top of the color maps and note regions of similarity with the branches~\cite{Pereira_2009, Franchini}. Using a bandwidth $J_{\rm eff}$ for the free-fermion dispersion (computed from $v_F$ of the qFL SF), we note a general matching with the three branches of the $Vt'_c$ model. Additionally, by kinematic constraints, the lower free-fermion branch remains the lower boundary of spectral support for all models away from half filling. Finally, using $J_{\rm bound}$ to fit the magnon dispersion, we note a qualitative agreement with the curvature of the upper branches for the net-attractive models, again suggestive of bound-state/excitonic physics due to an effective negative interaction~\cite{Franchini, Viswanath_1995, Pereira_2008}.

To further probe the high-energy regime, we study the lineshape of several momentum cuts by examining how $S_{\max}$ and the half width at half maximum, $\sigma_\pm = |\omega_\pm - \omega_{\max}|$ scale with $\gamma$, where $\omega_{\max, \pm}$ are defined as $S_q(\omega_{\max})= S_{\max}$ and $S_q(\omega_\pm) = S_{\max}/2$. Here, we perform the $\gamma$ and $\chi$ scaling limits concurrently through the ansatz $\gamma(\xi) = c/\xi(\chi)$, which adapts finite-size methods to uMPS. $c > 0$ is chosen to balance the spectrum resolution ($\sim\xi^{-1}$) with the line-shape relevant features~\cite{suppmat, Jeckelmann_2002}. Similar to the SF, we extract $\eta_{\max}$ as the slope of $\ln S_{\max}(\ln \gamma)$. From $\sigma_{\pm}(\gamma)$, slopes $\Sigma_{\pm} \sim 1$ characterize Lorentzians while $\Sigma_{\pm} \neq 1$ denote power laws. Together these quantities capture the qualitative changes in the lineshape with $q$.

Figure~\ref{fig:etaq} shows $\eta_{\max}(q)$, $\Sigma_{\pm}(q)$ for the attractive qFL: $VV_2$, $Vt_c$, $V_2t_c$. As remarked above, we hypothesize exciton-like physics in this dominant upper branch. This exciton originates from particle-hole symmetric two-particle excitations. Indeed, observe that $\Sigma_{\pm} \rightarrow 1$ for specific values of $q$ in the high-energy regime. This mirrors the line-shape evolution toward Lorentzian-like peaks, suggesting quasiparticle behavior in the upper branch. 
However, $\eta_{\rm max} \not\sim 1$. Its difference from unity is due to hybridization with the continuum, which decreases as the peak-continuum energy separation increases and/or the continuum spectral weight itself decreases. Indeed, the attractive-$tV$ model DSF possesses a fully disconnected bound state in the same momentum regime~\cite{Pereira_2008} yielding $\eta_{\max} \sim 1$, $\Sigma_{\pm} \sim 1$ (not shown). 
All this suggests that this exciton is a universal feature of attractive-qFL models.
In contrast, the absence of the exciton is a universal feature of repulsive-qFL models~\cite{suppmat} in line with the physics of the repulsive-$tV$ model~\cite{Franchini, Caux_2005a, Caux_2005b, Caux_2011}.

Our results show the qFL universality beyond the perturbative limit, lifting the need for exponentially precise resolution in experiments: Irrelevant interactions need not be small~\cite{Rozhkov_2014}. In platforms such as cold atoms and quantum simulators, interaction couplings can be made short range---as in our Hamiltonian class---and independently tunable and nullified~\cite{Trotzky2008, Das2024, Xu2025}. This selective control may enable stabilization of the qFL. Furthermore, momentum-resolved measurements---essential for probing qFL behavior---are standard in these setups.

\paragraph*{Conclusion.} We computed static and dynamic correlations for a class of lattice Hamiltonians realizing the universal features of qFL. Most significantly, these consist of a finite discontinuity in the momentum distribution, despite the presence of power-law singularities in the spectral function. Away from half filling, these models also display Landau quasiparticles. These disparate features shared across this class solidify the qFL as a robust phase of correlated 1D systems beyond the LL paradigm. Additionally, the dynamic structure factor reveals bound states near the upper threshold for attractive systems, while repulsive ones localize spectral weight within the two-particle continuum, with a renormalized bandwidth. {These FL-like characteristics of both ground state and excitations strongly suggest that the qFL paradigm allows for the restoration of FL physics in 1D systems.} Finally, preliminary results (not shown) indicate that including additional Hamiltonian terms (e.g., next-nearest-neighbor hopping) can also yield $K \sim 1$ regions in parameter space.

\begin{acknowledgments}
The authors thank A. Rozhkov for previous collaboration and fruitful discussions. A.E.F.\ and J.D.B.\ acknowledge support from the U.S. Department of Energy, Office of Basic Energy Sciences under Grant No.\ DE-SC0014407 (A.E.F.\ and J.D.B.). J.R.\ acknowledges support from the Office of the Vice-president of Research and Creative Activities and the Office of the Faculty of Science Vice-president of Research of Universidad de los Andes under the FAPA grant. Computational cluster research was conducted as part of a user project at the Center for Nanophase Materials Sciences (CNMS), which is a U.S.\ Department of Energy, Office of Science User Facility at Oak Ridge National Laboratory.
\end{acknowledgments}

\bibliography{references}

\begin{thebibliography}{52}
\bibitem [{foo()}]{footnote3}%
  \BibitemOpen
  \bibinfo {note} {The $Vt'_c$ model is equivalent to the $J_1$-$J_2$ Heisenberg model under a Jordan-Wigner transformation.}\BibitemShut {Stop}%
\end{thebibliography}

\end{document}


\title{Supplemental Material for \\``Universal quasi-Fermi liquid physics of one-dimensional interacting fermions"}

\author{Joshua D. Baktay}
\affiliation{Department of Physics, Northeastern University, Boston, Massachusetts 02115, USA}

\author{Adrian E. Feiguin}
\affiliation{Department of Physics, Northeastern University, Boston, Massachusetts 02115, USA}

\author{Juli\'an Rinc\'on}
\affiliation{Department of Physics, Universidad de los Andes, Bogot\'a, D.C. 111711, Colombia}

\date{\today}

\maketitle

\section{Introduction}

In the main text we characterized the excitations of quasi-Fermi liquid (qFL) systems. We observed its universal signatures in the spectral function (SF), as critical excitations in at least one sector whose exponents decrease with increasing momentum. We also showed that the dynamic structure factor (DSF) displays an upper branch for systems with an attractive effective interaction, associated with particle-hole excitations that are particle-hole symmetric. Here we present a complete characterization of the qFL ground state and how to stabilize it, following Ref.~\cite{Baktay_2023}. This demonstrates the other universal features of the qFL. Finally, we provide some supporting analysis for the DSF.

\section{Ground State Analysis}

\begin{figure}
    \centering
    \includegraphics*[width=0.48\textwidth]{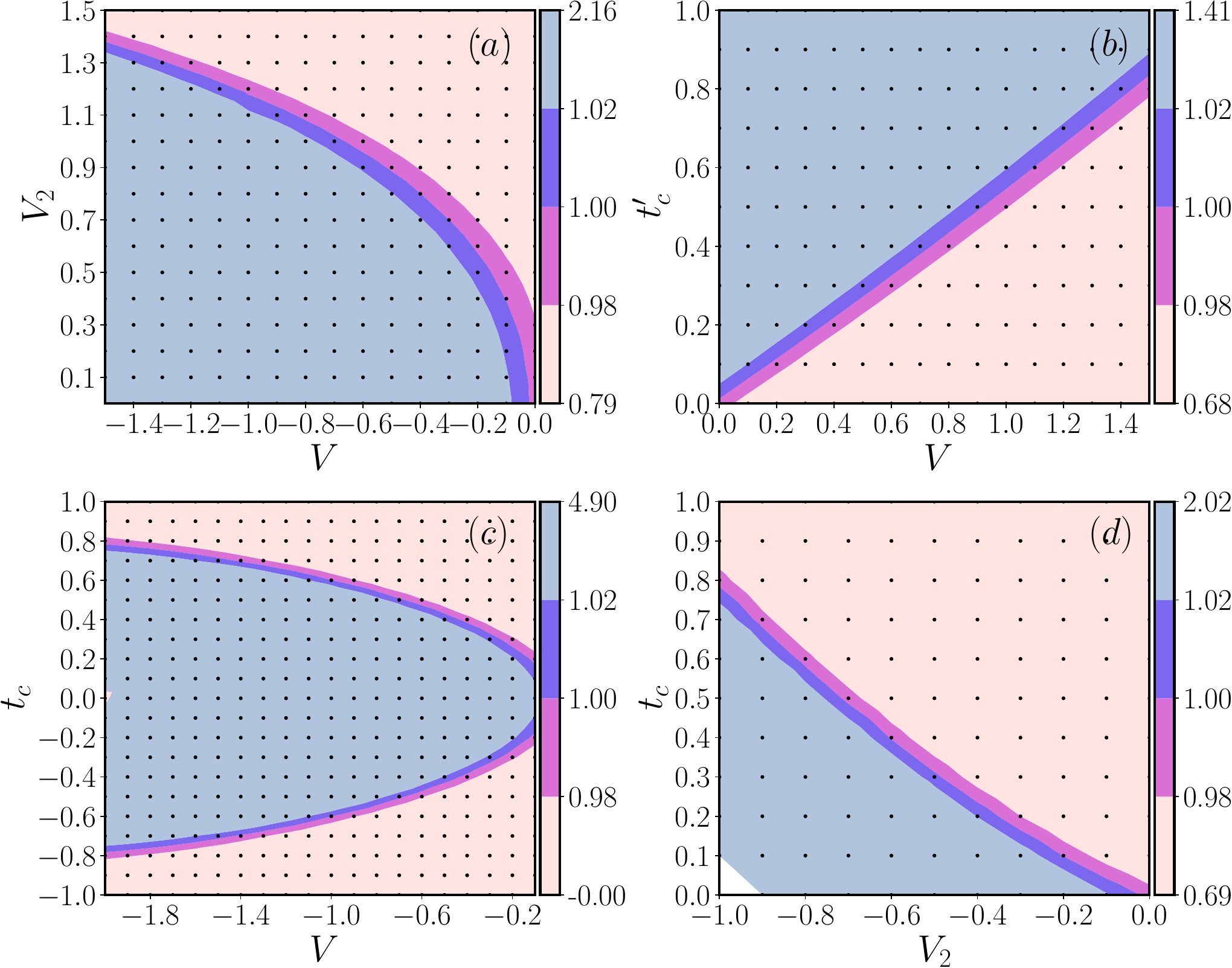}
    \caption{Phase diagram of the Luttinger parameter, $K$, for all four qFL models at (a) $\mu / t = 0$, (b) $\mu / t = 1.0$, (c) $\mu / t = 0$, (d) $\mu / t = 1.0$. The qFL is stable along each band where $0.98 \leq K \leq 1.02$. The relaxation of the strict $K=1$ constraint demonstrates the robustness of the qFL phase. For $K < 1~(K > 1)$ the ground state is dominated by charge-ordered (superconducting) fluctuations. In (c) we demonstrate the preservation of the qFL band under particle-hole exchange through its reflection about the $t_c=0$ axis (which also occurs for the $Vt'_c$ model---not shown). 
    From these bands we chose the following representative values for further study: $(V, t'_c) = (1.4, 0.8),~(V, V_2) = (-1.2, 1.2),~(V, t_c) = (-1.9, 0.8),~(V_2, t_c) = (-1, 0.8)$.}
    \label{fig:Kphase}
\end{figure}

\begin{figure}
    \centering
    \includegraphics*[width=.35\textwidth]{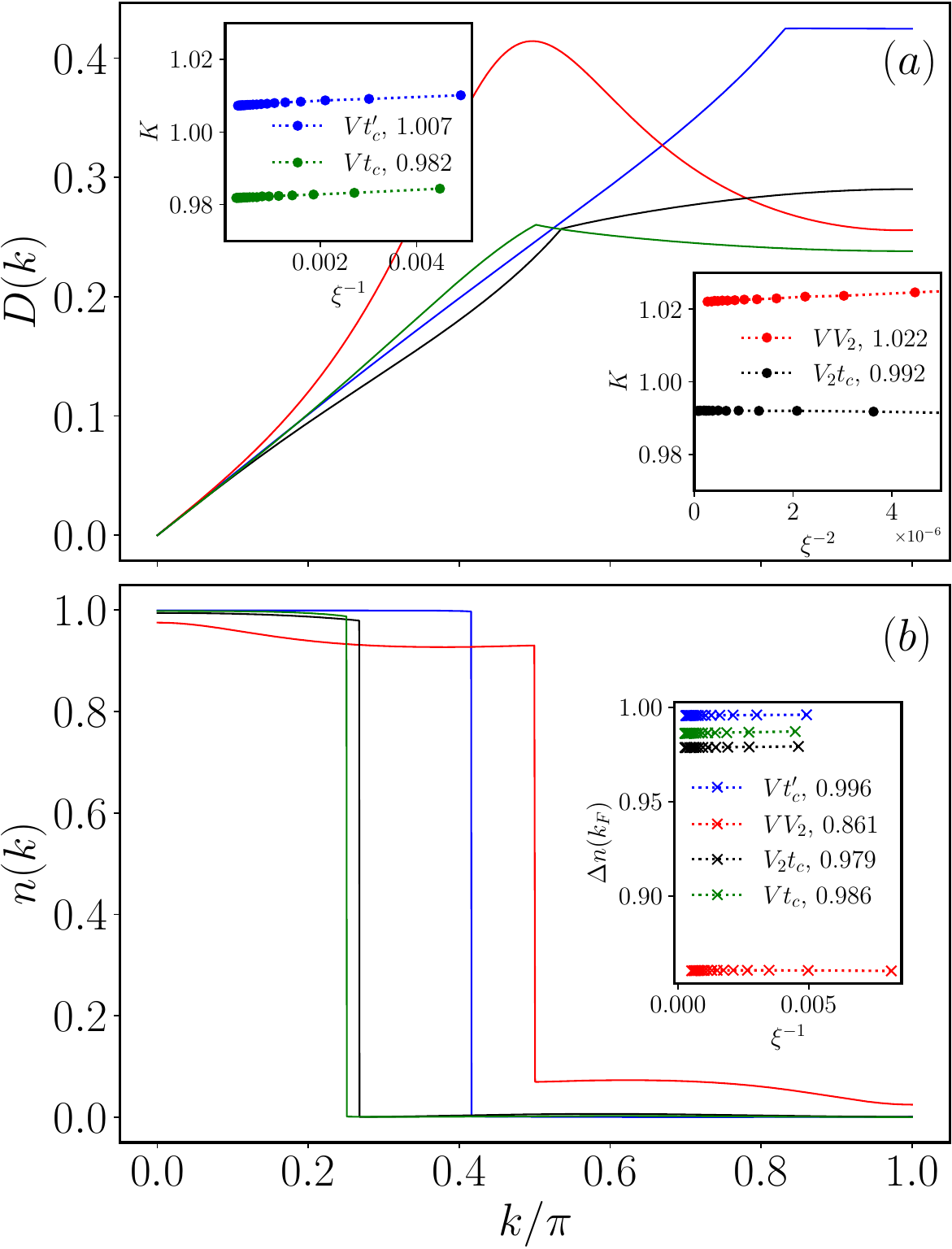}
    \caption{(a) Static structure factor $D(k)$ and (b) momentum distribution $n(k)$ for the representatives of all four qFL models. The insets show the FES analysis for the Luttinger parameter $K$ and the size of the discontinuity, $\Delta n(k_F)$. $K = 1.00(2)$ along with a finite $\Delta n(k_F)$ for $\chi \rightarrow \infty$ confirms the 
    qFL region in Fig.~\ref{fig:Kphase}. The different scaling exponents for $\xi$ define their relationship to $\chi$ and is in general model dependent.
    }
    \label{fig:scfs}
\end{figure}

We begin by analyzing the ground states of our qFL models, following Ref.~\cite{Baktay_2023}. According to the underlying Luttinger liquid (LL) field theory, the Luttinger parameter $K$ characterizes the strength (how strong) and nature (attractive, $K > 1$; repulsive, $K < 1$; noninteracting, $K = 1$) of the interactions. Setting $K = 1$ guarantees that at low enough energies the ground state is free-fermion-like with corrections at higher energies determined by the remnant irrelevant interactions. This is the defining feature of a qFL: it is perturbatively connected to a Fermi gas at low energies and stabilized by irrelevant interactions at high energies. By setting the Hamiltonian parameters of the class such that $K \simeq 1$, all our models exhibit a discontinuity in the momentum distribution at the Fermi momentum, $k_F$, as demonstrated below. This is characteristic of FL and stands in contrast to LL which manifest a universal power-law singularity at $k_F$.


It follows from LL theory that $K$ can be extracted as a low-momentum, linear approximation of the charge structure factor (CSF): the Fourier transform of the static density-density correlation function,
\begin{equation}
    D(k)=\sum_{j}\langle n_0 n_j\rangle e^{-ikj}.
\end{equation}
Using the uniform matrix product state (uMPS) approximation to the ground state to compute the expectation value $\langle n_0 n_j\rangle$, and therefore $D(k)$, $K$ can be extracted from its slope near $k \to 0^+$ as $dD(k)/dk=K/2\pi$~\cite{Giamarchi}. Thus, a judicious grid search was performed in Hamiltonian parameter space where $K$ was extracted from the CSF for each set of parameter values for each model of the qFL class. The results are shown in Fig.~\ref{fig:Kphase}. For each model we observe a continuous region of parameters where $K \simeq 1$. This represents a family of qFL candidates where the parameter values are tuned relative to each other to stabilize the state. We note that the regions surrounding the $K \simeq 1$ region is dominated by either charge-density-wave ($K < 1$) or superconducting ($K > 1$) instabilities, in agreement with previous literature. {In such regions, unlike the qFL, marginal interactions take over and dominate the low-energy physics.}

The contour of each region reveals an important point about the generality of the qFL physics. In none of our four models are the qFL parameters commensurate in value which one might expect given that such parameter adjusting serves to nullify marginal interactions. This illustrates that the qFL behavior is independent of a particular microscopic mechanism (like the intuitive competition between attractive and repulsive interactions) and instead depends purely on the effective couplings in the infrared limit.

The CSF of four representative candidates for each model is presented in Fig.~\ref{fig:scfs}(a). The smooth behavior indicates a metallic state, and the deviations at high momenta from the plateau-like trend of free fermions signify interactions. In the insets we show the finite-entanglement scaling of $K$ as a function of the uMPS correlation length $\xi$. This allows us to extrapolate our approximate value of $K$ to the infinite-$\chi$ limit corresponding to the exact value. With the high statistical correlation of the linear fit we are confident in our results.

Remarkably, note that $K$ for the $Vt_c$ and $VV_2$ models is not strictly 1, differing by $\approx 2\%$. This implies that strict calibration of Hamiltonian parameters is not necessary to realize the qFL, and may make an experimental realization more tractable. 

Next, we calculated the momentum distribution function,
\begin{equation}
    n(k)=\sum_{j}\langle c^\dagger_0 c_j \rangle e^{-ikj},
\end{equation}
similarly to the CSF~\cite{Vanderstraeten_2019}, which can be seen in Fig.~\ref{fig:scfs}(b). As was shown in Ref.~\cite{Rozhkov_2014}, the qFL ground state (again being perturbatively connected to free fermions) possesses a finite discontinuity in $n(k)$ at the Fermi point, $k_F$, in stark contrast to the kink power-law singularity of a LL. In order to show this we examined the finite-entanglement scaling (FES) of the size of the discontinuity about $k_F$. Specifically, we computed it as 
\begin{equation}
\Delta n(k_{F}) := n(k_{F} - \pi / \xi (\chi) )  - n(k_{F} + \pi / \xi (\chi) )
\end{equation}
where $\xi (\chi)$ is the correlation length computed from the uMPS transfer matrix as a function of the bond dimension, $\chi$. Given the inherently limited resolution of a finite-$\chi$ data set, a FES analysis was performed on $\Delta n(k_{F})$ for each model. This can be seen in the inset of Fig.~\ref{fig:scfs}(b), where $\Delta n(k_{F})$ clearly extrapolates to finite values as $\xi(\chi) \rightarrow \infty$.

The finite discontinuity in $n(k_F)$ along with $K=1$ as $\xi(\chi) \rightarrow \infty$ completes the qFL ground state characterization of our class of Hamiltonian models. {We note that the qFL is stable along the purple and magenta bands in Fig.~\ref{fig:Kphase}. The fact that the $K=1$ constraint need not be imposed shows the robustness of the qFL phase.}

\section{Entanglement scaling for dynamic correlation functions \label{FES}}

Let us discuss the extension of finite-entanglement scaling analysis to the realm of dynamic correlations computed using MPS, which we exploited to characterize the excitations of the qFL. The core idea of finite-entanglement scaling is based on the fact that the MPS correlation length, $\xi = \xi(\chi)$ diverges as $\chi \to \infty$, and that all observables obey scaling laws as a function of $\xi$~\cite{Tagliacozzo_2008, Pollman_2009}. This idea can also be posed in terms of the vanishment of the level spacing of the eigenvalues of the MPS transfer matrix as $\chi \to \infty$. This is fundamental when studying critical systems where no finite-length scales exist. Although the discussion uses $\xi$ as the scale parameterizing the FES for dynamic correlations, we remark that it is also possible to use other eigenvalues of the MPS transfer matrix and combinations thereof~\cite{Rams_2018, Vanhecke_2019}.

For concreteness, suppose we want to accurately compute the dynamic correlation function of operator $O_q$ with momentum $q$ at energy $\omega$,
\begin{equation}
    D_O(q, \omega) = \sum_\alpha \left| \langle {\rm 0} | O_q | \alpha \rangle \right|^2 \delta(\omega - \omega_\alpha(q)),
\end{equation}
where we have used the Lehmann representation. Here, $\ket{\rm 0}$ and $\ket{\alpha}$ stand for the ground state and excitations of energy $\omega_\alpha(q)$ with respect to the ground-state energy. The so-called spectral weight or form factor is defined by the overlap $W(\alpha, q) = | \langle {\rm 0} | O_q | \alpha \rangle |^2$. We can compute this quantity using MPS methods, such as the quasiparticle ansatz, which approximate excited states with finite entanglement and, hence, finite resolution $\gamma > 0$. The form of the dynamic correlation under such an approximation and exploiting translation invariance is
\begin{equation}
    D_O^{\xi, \gamma}(q, \omega) \approx \sum_\alpha W_\xi(\alpha, q) \frac{\gamma}{[\omega - \omega_{\alpha,\xi}(q)]^2 + \gamma^2},
    \label{convolution}
\end{equation}
where the excitation energy $\omega_{\alpha, \xi}(q)$ and the form factor $W_\xi(\alpha, q) = | \langle \Psi(A) | O_0 | \Phi_q^\alpha(B) \rangle |^2$ depend on $\xi$, again, because of the finiteness of $\chi$. The operator $O_0$ is now located at an arbitrary site 0. The ground and excited states have been approximated with a uMPS
$$
\left|\Psi(A)\right\rangle = \sum_{\{s\}} \left(\prod_{m \in \mathbb{Z}} A^{s_m}\right) |\{s\}\rangle,
$$
and a quasiparticle ansatz state~\cite{Haegeman_2012, Vanderstraeten_2019}
$$
\left|\Phi_q(B)\right\rangle = \sum_{n, \{s\}} \mathrm{e}^{i q n} \left[\prod_{m<n} A_L^{s_m}\right] B^{s_n} \left[\prod_{m>n} A_R^{s_m}\right] |\{s\}\rangle,
$$
where $A$ and $B$ are rank-three tensors, of dimension $\chi^2 d$, containing the variational parameters of the ground and excited states, respectively. $B$ represents a local perturbation on top of the uMPS ground state. In accordance with uMPS, we operate in the thermodynamic limit and thus the excitations have a well-defined momentum, $q$. 

The finite-entanglement scaling technique for dynamic correlation functions then consists of generating sequences of data for $D_O^{\xi, \gamma}$ for pairs $(\xi, \gamma)$, and studying its scaling behavior (how it changes as a function of $\xi$ and $\gamma$). It is expected that in the scaling limits $\xi \to \infty$ and $\gamma \to 0$ a construction of $D_O$ will be possible. Most generally the limit of $\xi$ must precede that of $\gamma$. Indeed, if $D_O$ possesses a continuous spectrum, then for finite $\xi$, equivalently finite $\chi$, if $\gamma \to 0$ the spectrum will be discrete and the lineshape will not be smooth. On the contrary, if the spectrum is discrete then we can safely evaluate the limit $\gamma \to 0$ first.

\begin{figure}
    \centering
    \includegraphics*[width=.35\textwidth]{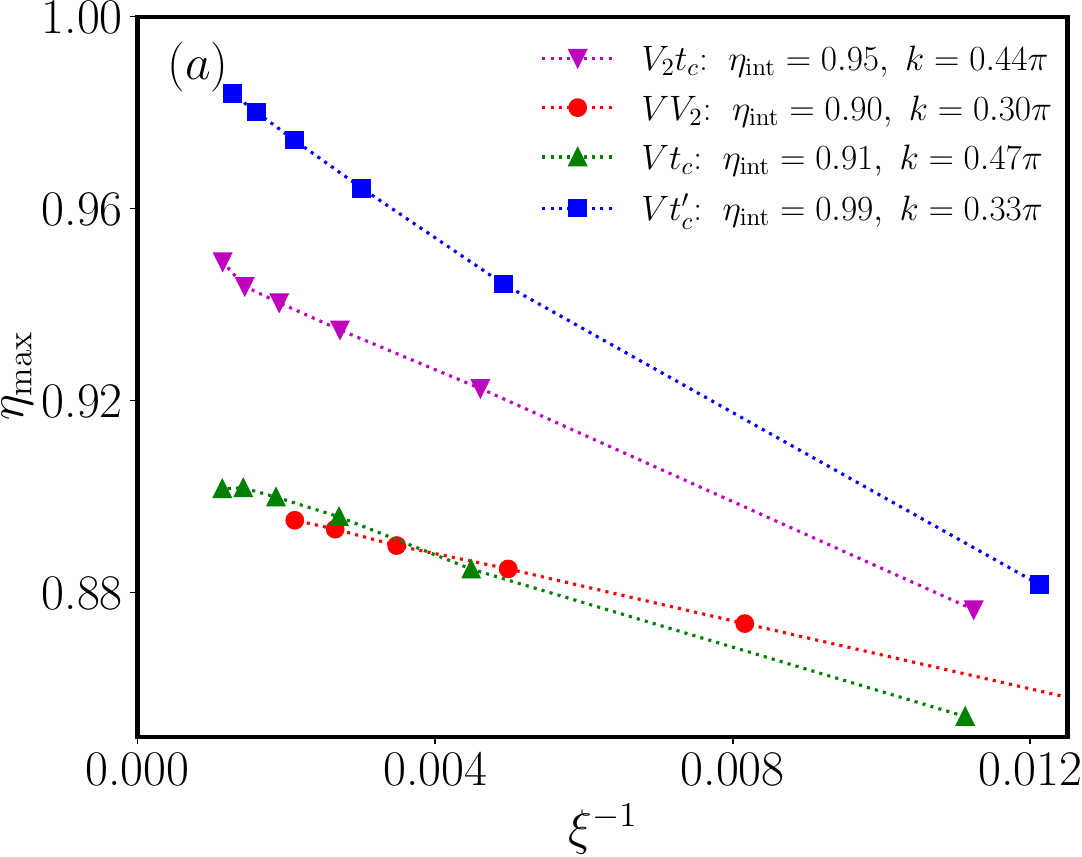} 
    \includegraphics*[width=.35\textwidth]{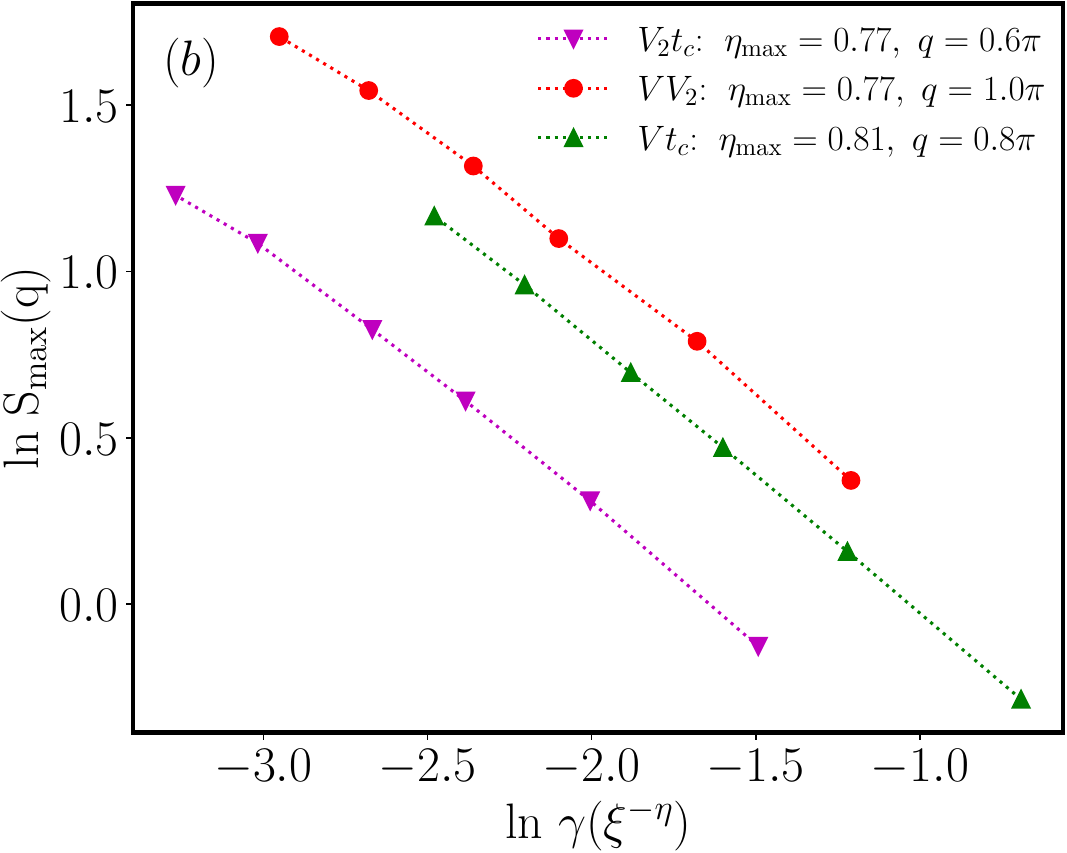}
    \caption{FES techniques for extracting scaling exponents according to Sec.~\ref{FES}. (a) Depicts the method used for $A(k, \omega)$ where a sequence of $\eta_{\rm max}$ is linearly fitted against $\xi^{-1}$ yielding the intercept $\eta_{\rm int}$ corresponding to $\eta_{\rm max}$ as $\chi \rightarrow \infty$. Each $\eta_{\rm max}$ is extracted by linearly fitting the peaks spectral weight against $\gamma$ on a log-log scale (not shown). (b) Depicts the method used for $S(q, \omega)$ where a sequence of $\xi$ concurrently gives a series of $\gamma$ that is linearly fitted against the peaks spectral weight yielding the slope, $\eta_{\rm max}$. In each plot we show for each model representative momentum values that span a wide energy range. The dotted lines are a guide to the eye. Each linear fit has a correlation coefficient $R\geq 0.99$.
    }
    \label{fig:dcf_fes}
\end{figure}

We require for the broadening $\gamma$ to always be larger than the largest level spacing, $\Delta \omega_{\alpha,\xi}$, of the excitation spectrum, i.e.~$\gamma > \Delta \omega_{\alpha,\xi}$ for fixed $\xi(\chi)$. Notice that as $\xi \to \infty$, $\Delta \omega_{\alpha,\xi} \to 0$, so the spectral response is smoothed out and closely resembles the exact result. Further requirements for $\gamma$ can be imposed depending on the scaling procedure utilized (see below).

The finite-entanglement scaling behavior of our MPS-represented dynamic correlations is first demonstrated by computing the $\gamma$ and $\xi$ limits not concurrently as follows (always maintaining $\gamma > \Delta \omega_{\alpha,\xi}$). It can be shown that the maximum, which occurs at $\omega = \omega_\alpha(q)$, of the convolution of $D_O$ with a Lorentzian diverges as $\gamma^{-\eta}$. The exponent satisfies $\eta = 1~(\eta < 1)$ for quasiparticle (power-law) excitations. Thus, for fixed $\xi(\chi)$, one can extract an approximate $\eta$ for a decreasing sequence of $\gamma$ values (an analogous technique was done in Ref.~\cite{Baktay_2023} albeit for fixed system size). Subsequently, the entanglement scaling occurs by generating a sequence of values for $\xi$ and extrapolating $\eta$ as a function of $\xi$ [see Fig.~\ref{fig:dcf_fes}(a)].

A more efficient and potentially more accurate approach is perform the $\gamma$ and $\xi$ limits concurrently by using a broadening that depends on $\chi$: $\gamma = \gamma(\chi) > 0$, and goes to zero as $\xi \to \infty$, which reflects the fact that a higher energy resolution is achieved as the excitations approximation is improved by increasing $\chi$. Leveraging previous work on finite-size scaling of dynamic correlation functions~\cite{Jeckelmann_2002}, we propose the broadening function
\begin{equation}
    \gamma(\xi) := \frac{c}{\xi}  \sim \frac{1}{\chi^\kappa},
\end{equation}
where $c > 0$ is a $\chi$-independent parameter freely chosen to ensure $\gamma(\xi) > \Delta \omega_{\alpha,\xi}$. Note that $\xi \to \infty$, as $\chi \to \infty$, so $\gamma(\xi) \to 0$, yielding an extrapolated estimation of $\eta$. In this alternative approach, generating a sequence of $\xi$ values automatically guarantees a sequence for $\gamma$ that ensures that the scaling limits are satisfied [see Fig.~\ref{fig:dcf_fes}(b)]. To extract the associated exponent $\eta_{\max}$ we follow the prescription outlined in the main text.

\begin{figure}
    \centering
    \includegraphics*[width=.35\textwidth]{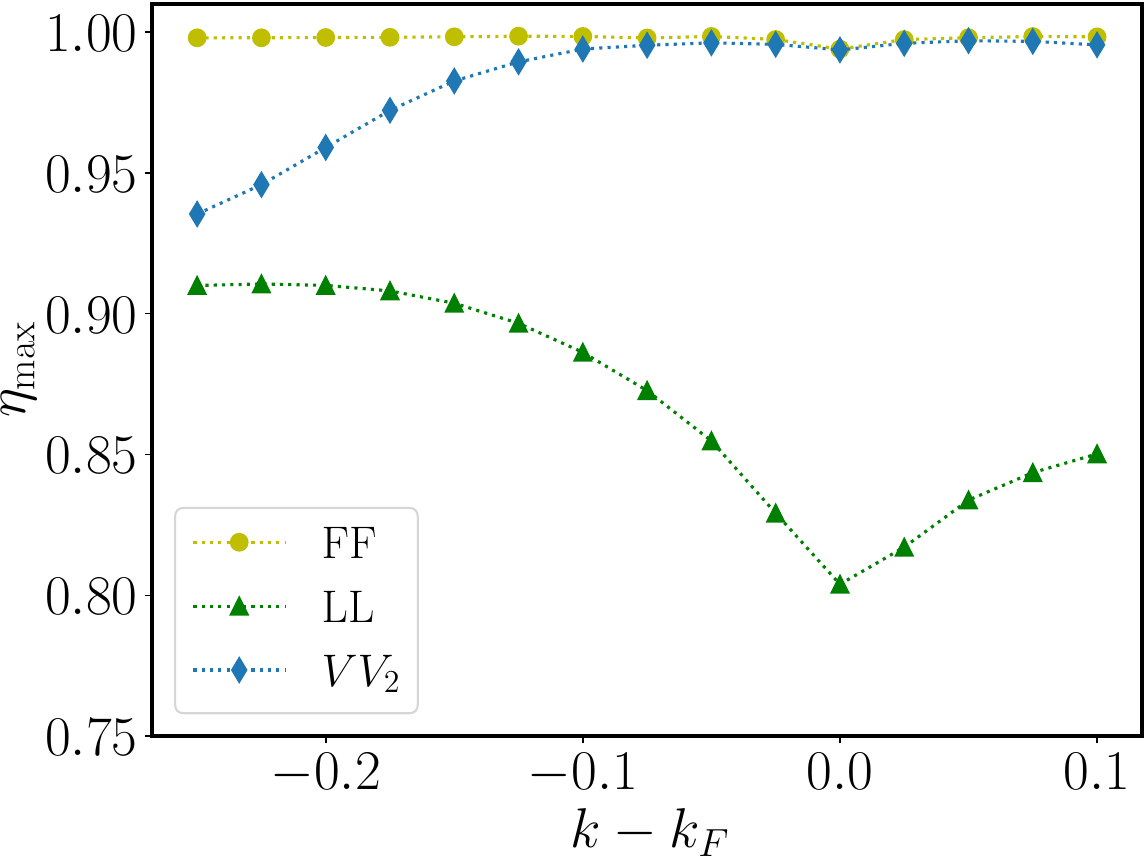}
    \caption{FES exponents as a function of momentum for the models presented in Ref.~\cite{Baktay_2023}. The FES analysis was performed as in the main text for bond dimensions up to $\chi=160$. The $VV_2$ qFL has a density of $n=0.3$ which breaks particle-hole symmetry and manifests the \emph{mixed} qFL signature: power-laws in only one sector, quasiparticles in the other. Contrary to the models in the main text, we find the collective excitations in the hole sector (see main text).
    }
    \label{fig:benchmark}
\end{figure}

\section{Excitations analysis}

\subsection{Spectral function results}

In Fig.~\ref{fig:benchmark} we reproduce the results of Ref.~[\onlinecite{Baktay_2023}] in order to benchmark the method described in the main text. The spectral function results of Ref.~[\onlinecite{Baktay_2023}] were computed for a finite-size system using time-dependent density matrix renormalization group (tDMRG), where only the $\gamma$ scaling was performed at fixed chain length. Thus our results improve on the previous in two ways: (1) they were calculated in the infinite-chain (thermodynamic) limit and (2) both the $\gamma$ and $\chi$ scaling limits were performed. 
The agreement between the two methods both confirms the validity of our method while also demonstrating that the finite-chain results survive in the infinite length limit. 

\subsection{Dynamic structure factor results}

\begin{figure}
    \centering
    \includegraphics*[width=.35\textwidth]{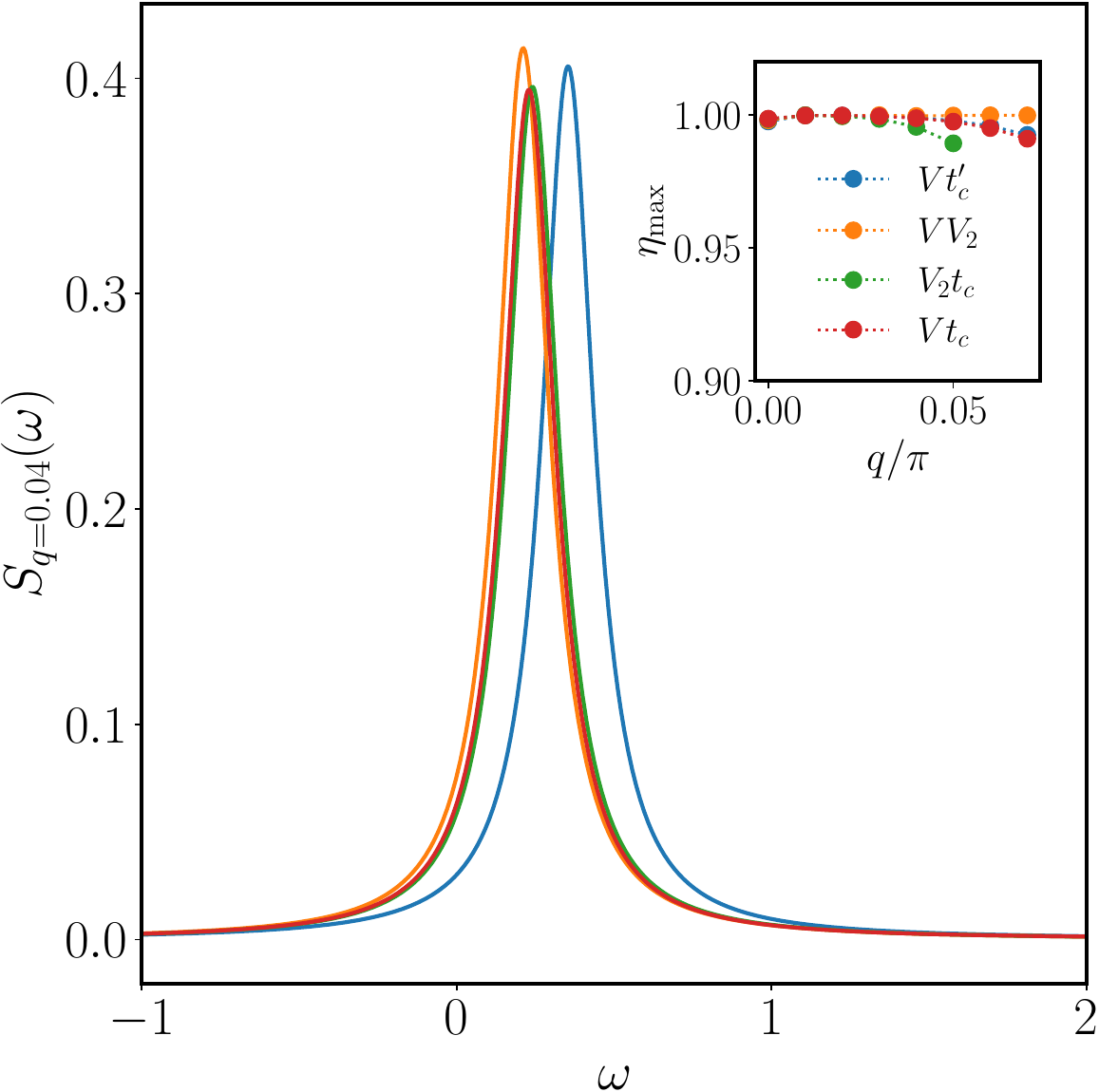}
    \caption{Momentum cuts of the dynamic structure factor at $q=0.04\pi \ll k_F$ for the qFL class of models presented in the main text at $\chi=104$ and artificial broadening $\gamma=0.1$. The Lorentzian lineshape confirmed by the scaling exponents in the inset confirms the presence of a sound mode for $q \ll k_F$.
    }
    \label{fig:soundmode}
\end{figure}

In Fig.~\ref{fig:soundmode} we present momentum cuts of the sound mode (low momentum region) for each Hamiltonian model of the qFL class, with fixed selected parameters, discussed in the main text. The clearly delta lineshape (which manifests in our calculations as a Lorentzian due to convolution) is confirmed by the inset where we show that the FES extrapolated scaling exponents $\eta_{\max} \sim 1$ for $q \ll k_F$. These exponents were computed by employing only the scaling limit for the broadening $\gamma$ at fixed $\chi$, or equivalently and in practice fixed $\xi$, which is sufficient for the purposes of our analysis. 

According to Refs.~\cite{Rozhkov_2005, Pereira_2007}, the presence of irrelevant operators like a nonlinear dispersion, or irrelevant interactions, introduces interactions between the LL bosons causing a quadratic broadening of the bandwidth in the DSF at low momentum that scales as: 
\begin{equation}
\delta\omega(q) = \omega_U(q) - \omega_L(q) = \frac{1}{\left| m^* \right|}q^2,
\end{equation} 
where $m^*$ is the effective mass, which can take positive and negative values, and $\omega_{L/U}(q)$ are the lower and upper boundaries of finite support in the DSF. 


\begin{figure}[b]
    \centering
    \includegraphics*[width=.35\textwidth]{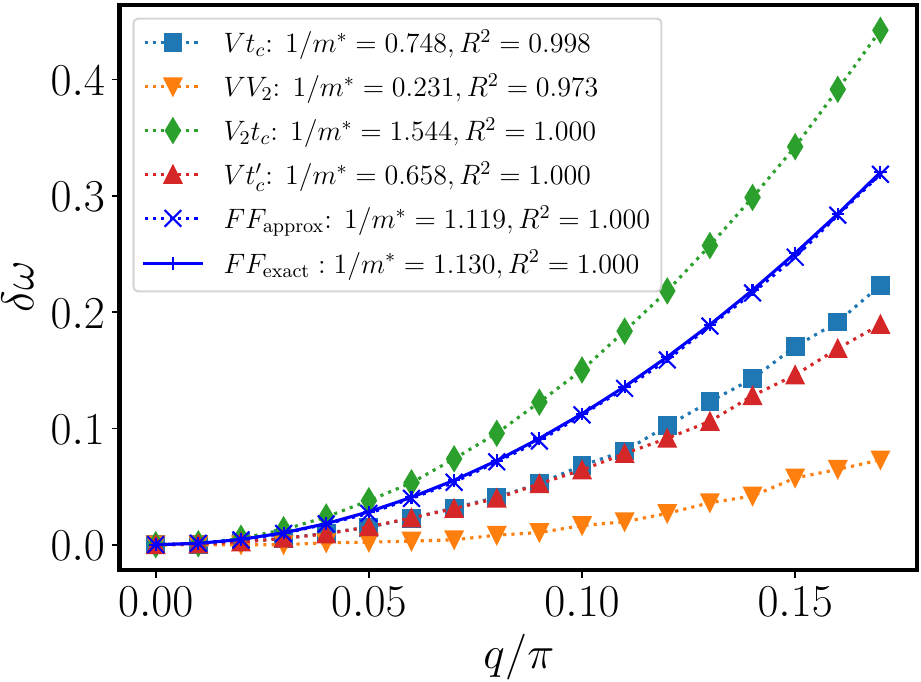}
    \caption{Quadratic fits of the bandwidth, $\delta\omega(q)$, as a function of momentum for the qFL models compared to free fermions at $\chi = 104$. For free fermions, we computed $\delta\omega(q)$ using both the exact dispersion and our simulations. The agreement between the two indicates the accuracy of our methods. For $q < k_F$ the bandwidth is significantly narrower relative to the continuum at higher momenta. Therefore, the density of excitations is much higher allowing for much more accurate results at the same bond dimension.
    }
    \label{fig:q2}
\end{figure}

In Fig.~\ref{fig:q2} we show $\delta\omega(q)$ for our qFL class of Hamiltonian models. The high correlation coefficients demonstrate that the DSF for our qFL all exhibit the same quadratic broadening, as expected from the general discussion above.

The broadening result for free fermions is $\delta\omega(q) \approx (J\cos{k_F})q^2$. Exploiting this result we can use our estimates of the effective mass $1 / \left| m^* \right|$ to predict the densities of our models and compare to the known densities: $n_{\rm predicted}=\{0.393, 0.219, 0.378, 0.462\},~n_{\rm known} = \{0.389, 0.268, 0.251, 0.500\}$ for the $Vt'_c$, $V_2t_c$, $Vt_c$, $VV_2$ models, respectively. From this we see a striking agreement for $Vt'_c$ compared to the other three. This indicates that the low-energy behavior of the $Vt'_c$ model is closer to a true free fermionic theory; the other three are more closely described by a free fermionic theory under the influence of irrelevant interactions, which we interpret as a Fermi liquid. This is reflected in the momentum distribution (see Fig.~\ref{fig:scfs}) where $Vt'_c$ is much closer to true step function with $\Delta n(k_F) \sim 1$.

\begin{figure}
    \centering
    \includegraphics*[width=.35\textwidth]{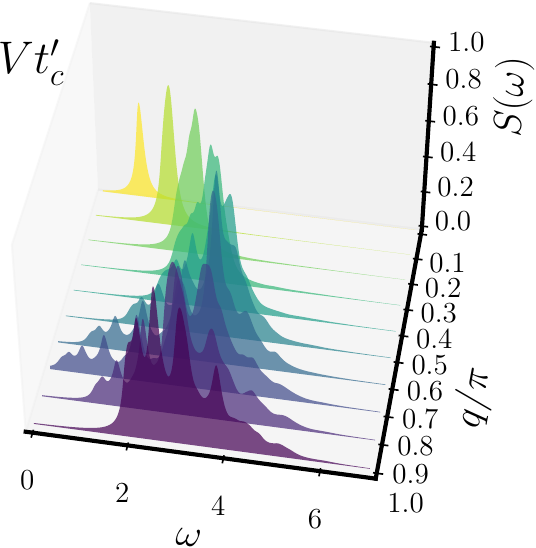}
    \caption{DSF lineshapes for different momentum values for the repulsive $Vt'_c$ qFL model. In contrast to the attractive qFL models discussed in the main text, as $q \rightarrow \pi$ the lineshapes accumulate spectral weight at intermediate energies.
    }
    \label{fig:projcuts_symVtc}
\end{figure}

Finally, in Fig.~\ref{fig:projcuts_symVtc} we show a series of lineshapes for the repulsive qFL, $Vt'_c$. Contrary to the attractive qFL models there is no high-energy bound state dispersion curve. Instead, the spectral weight accumulates in the middle of the continuum. The colormap in the main text suggests that this accumulation corresponds to a middle dispersion branch analogous to the repulsive $tV$ model.

\bibliography{references}